\documentclass[onecolumn, amsmath, amssymb, superscriptaddress]{revtex4}

\usepackage{dcolumn}
\usepackage{bm}
\usepackage{amsmath}
\usepackage{amsfonts}
\usepackage{mathrsfs}  
\usepackage{graphics}
\usepackage[dvips]{graphicx}
\usepackage{times}
\usepackage{setspace}
\usepackage{color}   
\usepackage{verbatim}
\usepackage[raggedright]{titlesec}
\usepackage[normalem]{ulem}
\usepackage{framed}
\usepackage{enumerate}
\usepackage{footmisc}
\usepackage[breaklinks=true, hyperfootnotes=false]{hyperref}
\usepackage{breakurl}

\usepackage[graphicx]{realboxes}
\usepackage{varwidth}
\usepackage{balance}
\usepackage{flushend}

\hypersetup{colorlinks=true,citecolor=blue, linkcolor=black,urlcolor=blue} 

\usepackage{hyperref}
\usepackage{amssymb}
\usepackage{mathbbol}
\usepackage{flushend}

\begin{document}
\title{Evolution of  biomedical  innovation quantified via\\  billions of distinct article-level MeSH  keyword combinations}
\author{Alexander Michael Petersen}
\affiliation{Department of Management of Complex Systems, Ernest and Julio Gallo Management Program, School of Engineering, University of California, Merced, California 95343, USA \\ apetersen3@ucmerced.edu}



\maketitle


{\bf \noindent 
To what degree has the vast space of  higher-order knowledge combinations   been explored and how has it evolved over time? To address these questions, we first develop a systematic approach to measuring  combinatorial innovation in the biomedical sciences based upon the comprehensive ontology of Medical Subject Headings (MeSH) developed and maintained by the US National Library of Medicine.  As such, this approach leverages an expert-defined knowledge ontology that features both breadth  (27,875 MeSH analyzed across 25 million articles indexed by PubMed that were published from 1902 onwards) and depth  (we differentiate between Major and Minor MeSH terms to identify differences in the knowledge network representation constructed from primary research topics only).
With this level of uniform resolution we  differentiate between three different modes of innovation contributing to the combinatorial knowledge network:  (i) conceptual innovation associated with the emergence of new concepts and entities (measured as the entry of new MeSH); and (ii) recombinant innovation, associated with the emergence of new combinations, which itself consists of two types: peripheral (i.e., combinations involving new knowledge) and core  (combinations comprised of pre-existing knowledge only).  
Another relevant question we seek to address is whether  examining triplet and quartet combinations, in addition to the more traditional dyadic or pairwise combinations, provide evidence of any new phenomena associated with higher-order  combinations.  
Analysis of the size, growth, and coverage of combinatorial innovation yield results that are largely independent of the combination order, thereby suggesting that the common dyadic approach is  sufficient to capture essential phenomena. Our main results are twofold: (a) despite the persistent addition of new MeSH terms, the network  is densifying over time meaning that scholars are increasingly exploring and realizing the vast space of all knowledge combinations; and (b) conceptual innovation is increasingly concentrated within single research articles, a harbinger of the recent paradigm shift towards convergence science. 

\begin{center}
{\bf combinatorial knowledge  $\vert$ knowledge network  $\vert$  conceptual innovation  $\vert$ recombinant innovation  $\vert$ PubMed  $\vert$ MeSH} 
\end{center}
 }

Scholars of the living world have long been faced with the monumental task of   systematically  cataloging the vast space of biological entities and subtypes.  
The emergence of computational and informatic capabilities has since  accelerated our ability to measure interactions within and between biological entities \cite{BMConvergence_2021}, such that the  task has  since shifted  towards  modeling   multi-scale  spatio-temporal complexity \cite{roehner2002patterns,bonaccorsi2008search,helbing2013globally}.
This new representation is manifestly a problem of combinatorics. Yet surveying the entire  frontier of that which is presently known -- what scholars have termed the  `adjacent possible' \cite{kauffman1995home,thurner2018introduction} -- is a compounding task. For this reason, many computational approaches   seek to leverage  vast  ontologies of codified knowledge and their relationships in  order to  automate  the survey of higher-order multi-component interactions. Examples  of this systematic approach to accelerating search and discovery include combinatorial exploration of  chemical compound ontologies to identify new drugs \cite{sadybekov2022synthon} and  novel thermoelectric materials \cite{tshitoyan2019unsupervised}.  

For the same reasons, combinatorial approaches to search and discovery may prove  valuable in exploring the vast solution space associated with many complex multi-dimensional challenges faced by society \cite{helbing2013globally,HBP_2020,arroyave2021social,ConvergenceMethods_2022}.
Against this backdrop, here we develop and apply a systematic ontology-based  approach for measuring  the size, growth and structure of combinatorial innovation in the biomedical sciences. 
This work contributes to prior  innovation research utilizing  article-level keywords that have either been  manually assigned \cite{Leahey_Sociological_2014,petersen2016triple,mishra2016quantifying,petersen2018mobility,BMConvergence_2021} or inferred by way of natural language processing of full or partial article text  \cite{gerlach2018network,tshitoyan2019unsupervised,palchykov2021network,arroyave2022NPS};  other notable approaches to the same end involve  identifying atypical citation combinations occurring within the reference lists of individual article \cite{Uzzi:2013}. 

In order to  avoid redundant and misspecified keywords, we leverage an existing and exogenously-defined  ontology of article-level keywords known as  Medical Subject Headings (MeSH) \cite{MESH}. As such, this work contributes to a growing literature operationalizing measures of innovation and knowledge networks  \cite{MESHMap,rodriguez2013quantifying,shi2015weaving,petersen2016triple,mishra2016quantifying,BMConvergence_2021} that take advantage of the clean structure and impressive historical coverage of research annotated by MeSH.
This multi-level biomedical ontology was  developed and continually maintained by the US National Library of Medicine \cite{MESH,MESHMap}, and is comprised of  more than 30,000 Medical Subject Headings (MeSH) used as article-level keywords   to classify  $\sim$30 million articles indexed within  PubMed. 
MeSH are organized in  a quasi-hierarchical ontology organized around 16  branch categories extending  up to 13 levels deep,  thereby defining different knowledge domains at various levels of specificity,  while further facilitating the definition of a metric distance between individual keywords \cite{BMConvergence_2021,HBP_2020,ConvergenceMethods_2022}. 
 
As in related work \cite{youn2015invention,mishra2016quantifying,Fortunato_2017_Science}, our main objective is to measure the size and marginal growth of recombinant innovation associated with the entry of new knowledge, operationalized here by tracking the first appearance of new MeSH terms and also higher-order MeSH combinations. 
To this end, Mishra and  Torvik   \cite{mishra2016quantifying} recently  analyzed the age of MeSH terms and MeSH-MeSH pairs occurring in research articles in order to develop quantitative measures of conceptual novelty. 
Continuing in this spirit,  we track all MeSH combinations  up to 4th order (i.e., tabulating unique quartets comprised of 4 MeSH combinations) according to the following four  objectives:
\begin{enumerate}
\item  develop a measurement framework for combinatorial knowledge production that supports  analyzing higher-order MeSH-MeSH (knowledge-knowledge) combinations;
\item  quantify the evolution of biomedical innovation  by systematically recording all unique MeSH combinations over time, in particular the first appearance of each. At the aggregate level, this integrative tabulation facilitates  defining the size,  growth, and  coverage of all possible combinations within the knowledge network;
\item  analyze the relative rates of  two principal modes of innovation  -- {\it conceptual} and {\it recombinant} -- by tracking the entry of new MeSH, in particular the degree to which new MeSH combinations correspond to two submodes of recombinant innovation -- {\it core} and {\it peripheral}.  To be specific, here {\it peripheral recombinant innovation} refers to  non-incremental innovation in  which  a  new MeSH combination also includes a new MeSH term, a  mechanism capturing combinatorial innovation at the knowledge frontier.  The complementary scenario, {\it core recombinant innovation},  refers to new combinations constructed from pre-existing entities only, and is a proxy for more integrative   refinements to the knowledge network. 
\item  account for variation in the  significance of article descriptors  by  distinguishing between primary  and secondary keywords.
\end{enumerate}
As in related work \cite{BMConvergence_2021}, this last objective manifests as   parallel analyses, one based upon ``All'' MeSH, and the second focusing only on ``Major'' MeSH which capture just the primary research topics, which thereby facilitates insightful juxtaposition.

\vspace{-0.3in}
\section*{Background}
\vspace{-0.2in}
\noindent This work contributes to the literature on combinatorial innovation, which has been developed in  several research streams: from theoretical approaches in economics to specify the knowledge production function as it relates to economic growth \cite{weitzman1998recombinant}; to non-equilibrium statistical physics models of  evolutionary processes based upon combinatorial interactions  \cite{thurner2018introduction}; and empirical research on the evolution of industrial innovation based upon analysis of the frequency of International Patents Classification (IPC) categories  \cite{fleming2001recombinant,youn2015invention,napolitano2018technology}.

Approaches to quantifying the growth of knowledge production  use various methods to define the space of entities and their combinations, which together serve  as a proxy for recombinant innovation. 
For example,  scholars have sought to measure the number of inventive classes and their distinct inventive combinations, showing that  the number of distinct combinations increases proportion to the number of new patents, meaning that the amount of new combinations per new patent is roughly constant  \cite{youn2015invention}.
Supporting evidence in the academic domain, based upon research spanning all fields of science, finds that the number of unique phrases in research article titles (a proxy for knowledge production based upon the total size of the topic space) also  follows a linear growth pattern, despite the volume of scientific research production (measured as the total number of research articles published per year) following exponential growth  pattern  \cite{Fortunato_2017_Science} $^{a}$.  
\footnotetext[1]{The details of how knowledge and its proxies are defined are likely to affect the assessment of innovation and its dynamics. 
 For example,  measuring the space of entities by tokenizing natural language  \cite{Fortunato_2017_Science}  adopts an endogenous definition of the concept space, since authors independently construct  titles from  select words that largely reflect disciplinary and other contextual factors. Instead,  exogenous constructions of the concept space are likely to be more uniform, and thereby avoid the challenges of accounting for stylistic and semantic aspects of language evolution. However, the tradeoff to an exogenous classification is the effort required to systematically tag all research articles, either manually or automatically. In the present case, individual MeSH terms are   manually assigned to research articles by expert annotators at the US National Library of Medicine. }
From this perspective, the dichotomy of exponential growth of production and linear growth of innovation \cite{youn2015invention,Fortunato_2017_Science} suggests that the  knowledge network -- comprised of  entities and their relationships that are codified and accepted by communities of scholars --  evolves by way of densification, as opposed to expansion at its surface deriving from the addition of new concepts -- which is a research question that the present work seeks to address.
 
Another consideration are the  drivers of  change. The exponential growth of the researcher population combined with  invaluable productivity innovations (e.g. computer-aided word processing and the  digitization of  journals)  together largely   explain the exponential trends in scientific production \cite{petersen_citationinflation_2018}. 
Drivers of innovation are less well-understood, as they more acutely depend on institutional and behavioral factors. The propensity for researchers to integrate  existing  knowledge, as opposed to exploring  new knowledge and knowledge-knowledge combinations, is largely affected by the risks associated with exploration \cite{fleming2001recombinant,Uzzi:2013}.
One should also consider the practical limits  that define the situational objectives and outcomes of knowledge producing activities. A research  project is  typically  focused around a few specific research questions grounded in prior research, which may  explain why contributions to innovation by individual articles appears to be  incremental (linear) \cite{youn2015invention,Fortunato_2017_Science}.
Another consideration is the compounding effort associated with combinatorial integration of new knowledge. That is, for each incremental advancement introducing one new piece of  knowledge  to a knowledge network of size $N$, then there are in principal $N-1$ new  pairwise combinations that a scholar would   want to  consider if the objective is to be exhaustive. 
Indeed, the process of integrating existing knowledge involves signifiant levels of uncertainty, as not all combinations are insightful or useful; and in practice, there is also higher variability in the potential value of the combinations that involve new entities, as illustrated in research on patent citations conditioned on how inventors integrate unfamiliar components and   their combinations  \cite{fleming2001recombinant}.

\begin{figure}
\centering{\includegraphics[width=0.85\textwidth]{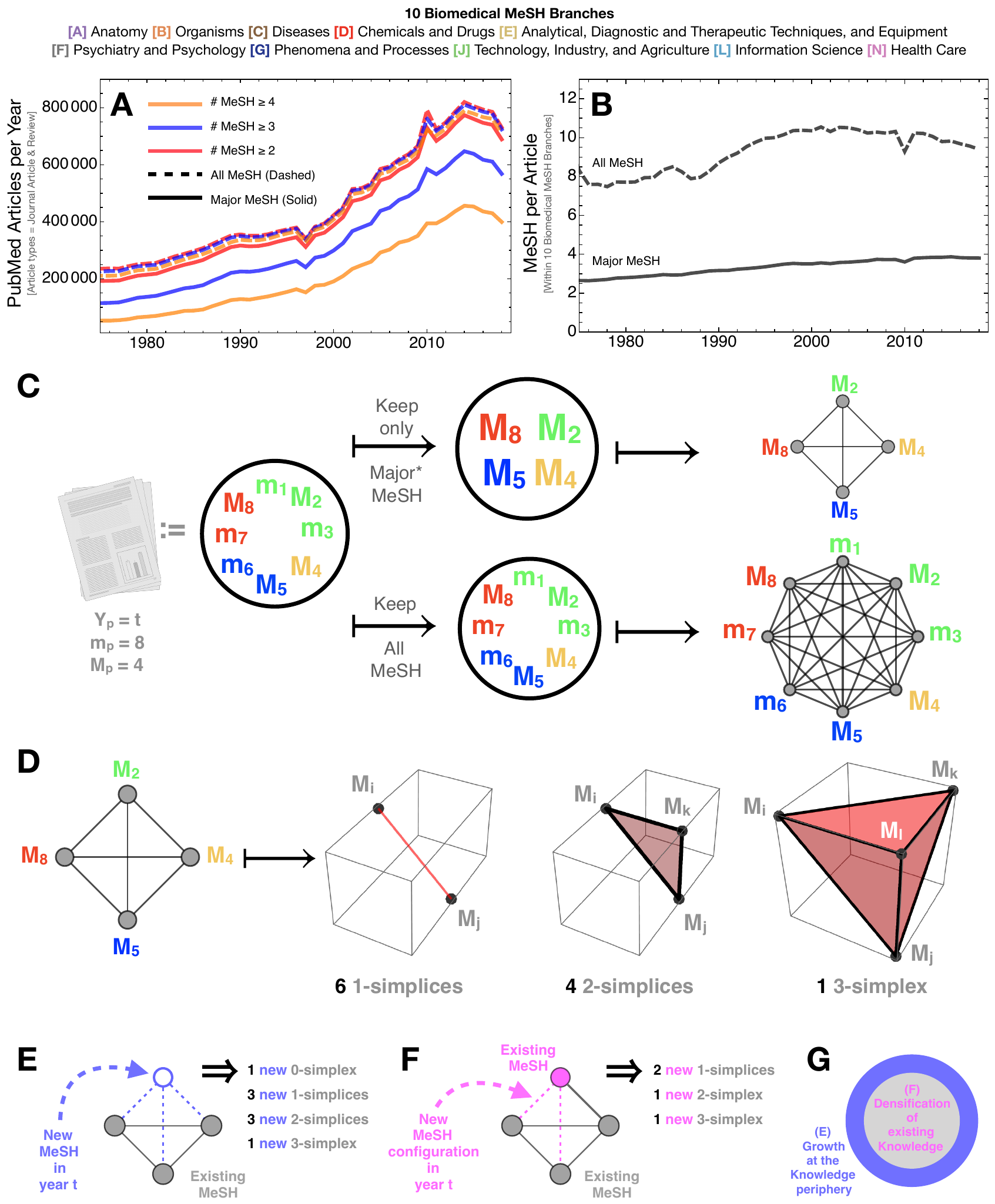}}
\caption{ \label{fig1}    {\bf Framework for quantifying the growth of combinatorial  knowledge associated with  conceptual and recombinant innovation.} We used the vast ontology of Medical Subject Headings (MeSH) developed and systematically implemented by the U. S. National Library of Medicine within the PubMed research article index. Owing to our focus on measuring recombinant innovation, we only analyzed articles featuring at least two MeSH belonging to at least one of the following biomedical-oriented MeSH branches: A, B, C, D, E, F, G, J, L, N. The six branches excluded from our analysis (H,I,K,M,V,Z) tend to focus around non-biomedical concepts and entities such as geographic locations and disciplinary contexts; see \cite{BMConvergence_2021} for more details regarding the MeSH knowledge network ontology as well as our select biomedical branch  criteria.
{\bf (A)} Number of research articles from PubMed index analyzed by year; we only analyzed article types  classified in PubMed as ``Journal Article'' or ``Review'' and excluded other content types such as commentaries, editorials and the like. {\bf (B)} The average number of  MeSH terms per article counted in two ways: (i) counting ``All'' MeSH terms (dashed line); (ii) counting only ``Major'' MeSH terms indicated by an $(^{*})$ in PubMed records (solid line). {\bf (C)} An illustrative article published in year $Y_{p}$ that features eight total MeSH ($m_{p}$=8), with four of those being Major category descriptors ($M_{p}=4$), which represent the article's principal  research topics.  {\bf (D)} From the resulting set of MeSH (either All or just Major) we then calculate all combinations  of $k+1$ MeSH  represents a unique k-Simplex. {\bf (E-F)} Distinguishing between two modes of innovation. {\bf (E)} {\it Conceptual innovation}: the entry of new MeSH terms representing new concepts, entities and existing entity sub-types is one  source for new  k-Simplices. {\bf (F)} {\it Recombinant innovation}: The reconfiguration of pre-existing MeSH into novel combinations is a second source for new k-Simplices. {\bf (G)} Conceptual (recombinant) innovation tends to occur at the periphery (core)  of the knowledge network.}
\end{figure}

\vspace{-0.3 in}
\section*{Methodology}
 \vspace{-0.2in}
\subsection*{Motivation} 
 \vspace{-0.2in}
\noindent We build on recent work analyzing higher-order  multi-entity relationships,   extending beyond the dyadic network  framework whereby at most two fundamental entities or nodes  are connected by a link,  to  a higher-order representation based on a simplicial set, whereby connectivity among multiple nodes is represented by a hyperedge    \cite{battiston2020networks,battiston2021physics}. Such a generalization is in principle a powerful elaboration; however, depending on the underlying processes giving rise to multi-entity combinations, such higher-order representation may not necessarily yield new insights beyond the information contained in the traditional pairwise-interaction network representation  \cite{merchan2016sufficiency}. 

Indeed, it could be that new phenomena identified by higher-order representations are redundant, or otherwise  inconsequential, given that they may be increasingly rare, with implications at the margins of what is measurable or inferable. By way of   analogy, consider the diagramatic framework for tabulating all contributions to  particle-particle scattering amplitudes championed by physics Nobelist Richard Feynman, which brilliantly reduces to a multiplicity of fine-structure constants $\alpha = 1/137$, such that each Feynman diagram vertex contributes a relative likelihood factor of $\sqrt{\alpha}$ \cite{feynman1949space}. Hence, increasingly elaborate Feynman diagrams involving higher-order particle interactions are also significantly less likely to occur, and hence contributes just marginally to the overall likelihood of particle-particle interaction. Such could also be the case for the information captured in higher-order knowledge  combinations, and is one of the motivations for  the systematic approach developed in this study.

More specifically, in order to explore whether any ``new physics'' emerges when accounting for higher orders of combinatorial innovation, we systematically tabulated all combinations of article-level keywords  through the fourth order -- i.e., tracing all distinct k-simplices of order $k=1$ (i.e. MeSH pairs), $k=2$ (MeSH triads) and $k=3$ (MeSH quartets) --  over time as they emerge in the biomedical literature. In this way, we are primarily using the simplex framework to facilitate exact counting of unique keyword combinations, and as such in this work we do not analyze the aggregate simplicial complex comprised of the union of all simplices into a higher-order network.$^{b}$
\footnotetext[2]{ This strategy is not just practical, but also reasonable, considering that the average PubMed article is tagged by numerous MeSH keywords -- see {\bf Fig. \ref{fig1}}(B). 
At this resolution level, the set of $m_{p}$ MeSH, on average ranging between 8 to 10 per article, are sufficient to capture both core and peripheral concepts defining the research. And so the value of one additional MeSH term is marginal at this level of detail. Yet, if we were to consider two articles, one with $m_{p}$ MeSH and another with $m_{p}+1$ MeSH, where all MeSH in the first article belong to the second article, then the articles differ in just the one additional MeSH term. It seems unreasonable to consider these articles as completely distinct, which would be the case if for each article we only tabulated the single ($m_{p}$-1)-simplicial complex formed by all the MeSH. Instead, we systematically decompose an article into all k-simplex combinations sampled from the set of $m_{p}$ MeSH, which is also robust to the time-dependent variation in the average number of MeSH per article over time.}

Systematic assessment of the  number of distinct knowledge combinations takes explicit advantage of the ontological features of the MeSH system, which catalogs unique concepts and entities by way of its thesaurus-like design. In this way, the MeSH ontology overcomes the limitations of  entity representations that suffer from redundancy (two different terms that represent the same concept) and simplicity (terms that are too shallow in their definition, such as in the case of broad category systems). These problems tend to emerge when the system of entities are not homogenized, reflecting variation in authors, disciplines or other linguistic features of their description that manifests as a big challenge when defining topic categories from raw text. In this way, the MeSH ontology reduces the  degrees of freedom in the representation of biomedical knowledge, while at the same time accounting for the vast variability in research topic breadth and depth. For example, ``Telomere Homeostasis'' [MeSH Unique ID D059505] which is synonymous with other entry terms  ``Telomere Length Maintenance'' and ``Telomere Lengthening'', yet is distinct from ``Telomere Shortening'' [MeSH Unique ID D059506], which is a distinct concept relating to the process of Telomere growth as opposed to the ``Telomere'' entity itself, which is an altogether distinct MeSH term [D016615]. 

In summary, the hierarchical structure of the MeSH knowledge tree  provides  adequate breadth, specificity, and uniformity to systematically perform historical analysis of  combinatorial innovation at  high conceptual resolution  \cite{petersen2016triple,BMConvergence_2021}.  In particular, since the ontology is controlled and maintained by a select unit at the National Library of Medicine, which has back-catalogued articles to the early 20th century (the first article with MeSH is from 1902), then the first appearance of a given MeSH or MeSH simplex can be accurately recorded and tabulated. In what follows, we focus on comparing trends observed in the modern era of biomedical research (1975-present) since the coverage of PubMed vastly expanded in the post-war era, and also because the number of  MeSH per article approached present levels since around 1975 \cite{mishra2016quantifying,petersen2016triple}.$^{c}$ 
\footnotetext[3]{While we only present results for 1975 and onwards, it's important to note that we started tabulating the first occurrence of each realized simplex from  1902 and onwards to avoid left-censoring bias  of cumulative tallies.
}

 \vspace{-0.2in}
\subsection*{Data} 
 \vspace{-0.2in}
\noindent We  represent the combinatorial  knowledge network as it manifests at the  article-level in the form of MeSH keywords, which belong to  controlled vocabulary of scientific concepts and entities organized in a  quasi-hierarchical relational tree. The official MeSH tree is maintained by the U. S. National Library of Medicine and  is comprised  of 16 top-level category branches: A, B, C, D, E, F, G, H, I, J, K, L, M, N, V, Z (see   https://meshb.nlm.nih.gov/treeView). Following previous work focusing on the biomedical and health science branches  \cite{BMConvergence_2021}, here we also focus on 10 branches: Anatomy [A], Organisms [B], Diseases [C], Chemicals and Drugs [D], Analytical, Diagnostic and Therapeutic Techniques, and Equipment [E], Psychiatry and Psychology [F], Phenomena and Processes [G], Technology, Industry, and Agriculture [J], Information Science [L], Health Care [N].  These branches are comprised of 27,875 individual MeSH terms, providing a comprehensive and detailed representation of  biomedical knowledge.
At the same time we ignore non-technical  MeSH that are tangential to biomedical innovation -- specifically, 1763 MeSH terms   belonging to the following 6 branches:   Disciplines and Occupations [H],  Anthropology, Education, Sociology, and Social Phenomena [I], Humanities [K], Named Groups [M], Publication Characteristics [V] and Geographicals [Z]. 

We downloaded all articles indexed by PubMed in  2020. In what follows we present results  based upon the subset of  research-oriented content indexed by PubMed defined by the following criteria:  (a) publications classified as ``Journal Article'' or ``Review''; (b)  annotated by 2 or more MeSH terms, resulting in a sample of $\sim$25 million articles. {\bf Figure \ref{fig1}}(A) shows the total number of articles by year, distinguishing between articles with 2, 3 and 4+ MeSH terms. To account for variability in the weight associated with an article keyword, we leveraged additional annotation information that identify the  ``Major MeSH'' terms, a subset of the entire set of keywords that  represent the primary research topics. This two-level annotation system is implemented in PubMed by way of an asterisk (*) next to those MeSH that are distinguished as Major. Hence, in what follows we perform and compare calculations based upon the entire set of MeSH annotations (denoted by ``All'') versus just the Major MeSH subset (denoted by ``Major''). 

 {\bf Figure \ref{fig1}}(B) shows that the average number of Major MeSH belonging to the focal 10 branches has increased from around 2.5 to 4 over the period 1975-2018; whereas, the average total number of MeSH has increased from 8 to roughly 10 over the same period. Hence, this refinement significantly reduces the number of MeSH per article considered in the Major representation. By way of example, not all articles will have sufficient numbers of Major MeSH to contribute to our analysis of quartets,  which requires there to be at least $4$ Major MeSH;  hence, we cannot include those research articles with just 1, 2 or 3 Major MeSH in our analysis of quartets. Consequently, we must also account for the variable sample sizes by year, depending on whether a sample includes articles with 2+ or 3+ or 4+ keywords. As such, we denote the set of articles having $k$ or more MeSH terms by $P_{k}$.  
 
\vspace{-0.2in}
\subsection*{Measures and Notation} 
\vspace{-0.2in}
\noindent  {\bf Figure  \ref{fig1}}(C) illustrates the process for  counting simplices of order k based upon the set of MeSH annotating a given publication $p$ published in year $Y_{p}$. The set of $M_{p}$ Major MeSH terms is a  subset of the  full set of $m_{p}$ MeSH terms. From these two MeSH sets we systematically tabulate all ${M_{p} \choose k}$ (respectively, ${m_{p} \choose k}$)  combinations for a given simplex order $k$. For example, {\bf Fig.  \ref{fig1}}(D) shows the number  ${M_{p} \choose k}$ of unique k-simplices derived from a set of four Major MeSH terms for $k=$ 1, 2, 3. For a given set of articles  we repeat this tabulation procedure, which yields a set $S_{k}$ of unique k-simplices, e.g. tabulated either within a specific year $t$ or aggregating all articles thru that year $t$.  In the latter scenario, we denote the cumulative number of distinct simplices of order $k$ thru year $t$ as $C_{t}(S_{k})$.  Consequently,  the number of new simplices emerging in $t$ is given by $\Delta C_{t}(S_{k}) = C_{t}(S_{k})-  C_{t-1}(S_{k})$. Similarly, the number of unique MeSH   tabulated across a particular set of articles is $N_{m}$; we represent the cumulative number of distinct MeSH  thru year $t$ as $C_{t}(N_{m}) \equiv C_{t}(S_{0})$, where the last equality follows since individual MeSH terms are also 0th order simplices;  the number of new MeSH appearing in year $t$ is  $\Delta C_{t}(N_{m}) = C_{t}(N_{m}) - C_{t-1}(N_{m}) $.  

Given a set of $C_{t}(N_{m})$ unique MeSH realized thru year $t$, the total number of possible k-simplices is given by ${C_{t}(N_{m}) \choose k}$. Hence, we can  measure the exact fraction of all possible k-simplices  appearing thru year $t$ as $F_{t}(S_{k})$ = $C_{t}(S_{k}) /  {C_{t}(N_{m}) \choose k}$. One limitation to our counting method is that we neglect simplex frequencies. Consequently, the tally $C_{t}(S_{k})$ may include many spurious  k-simplices that occurred just once. A potential future avenue of research would develop a measure that incorporates a counting weight that is proportional to the k-simplex frequency, or that implements a counting threshold to eliminate spurious k-simplices that only occurred once.

This counting framework facilitates systematically measuring two modes of innovation  -- {\it conceptual} and {\it recombinant}. The rate of conceptual innovation is   measured by $\Delta C_{t}(N_{m})$.  Similarly, the rate of recombinant innovation is  measured by $\Delta C_{t}(S_{k})$. Moreover,  
we can also specify what fraction of $\Delta C_{t}(S_{k})$ involve new MeSH appearing for the first time in the same year, corresponding to   {\it peripheral recombinant innovation}, and its complement  {\it core recombinant innovation} referring to new $S_{k}$ that are only comprised of pre-existing MeSH.

\begin{figure}
\centering{\includegraphics[width=0.99\textwidth]{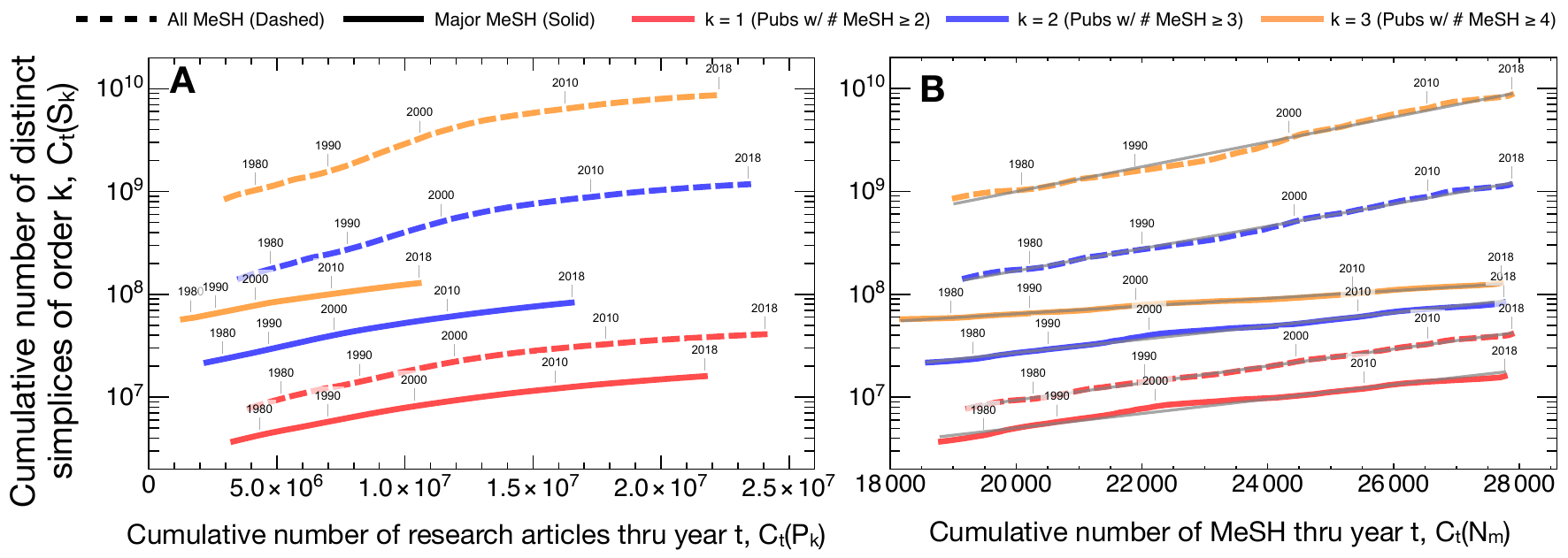}}
\caption{  \label{fig2}    {\bf Growth of the combinatorial knowledge network.} 
The set of   MeSH combinations realized in published biomedical literature   reveals the explored portion of the  knowledge network. We operationalize measuring this  revealed knowledge network by tabulating the cumulative number of distinct simplices of order $k$, measuring the 
size of the knowledge network as it grows in relation to (A) the total volume of research published, and (B) the total number of MeSH entities (corresponding to vertices, or simplices of order $k$=0).  By way of example, $C_{t}(S_{1})$ counts the total number of realized MeSH-MeSH dyads  (links) comprising the first-order knowledge network analyzed in ref. \cite{BMConvergence_2021}.  {\bf (A)} Common trends observed for the cumulative number of distinct k-Simplices $C_{t}(S_{k})$ as a function of  total volume of articles published, $C_{t}(P_{k})$ --  with a decreasing slope appearing in the early 2000s, corresponding to a decreasing rate of new simplices per publication relative to the earlier period. {\bf (B)} Persistent exponential  growth of $C_{t}(S_{k})$ with total size of the MeSH ontology. Each solid gray line represents the best exponential model fit.  }
\end{figure}

 \vspace{-0.3in}
\section*{Results}
 \vspace{-0.2in}
\subsection*{Growth of the biomedical knowledge network}
\vspace{-0.2in}
\subsubsection*{Combinatorial innovation and scientific production}  
\vspace{-0.2in}
\noindent {\bf Figure \ref{fig2}}(A)  shows the total number of MeSH simplices of order $k$, denoted by $C_{t}(S_{k})$, as function of the cumulative number of research articles indexed by PubMed, denoted by $C_{t}(P_{k})$.  There are six curves, because for each simplex order $k=1,2,3$ we calculated $C_{t}(S_{k})$ and  $C_{t}(P_{k})$ for each of the two MeSH refinements (All, Major).  Since there needs to be at least $k$ MeSH for an article to contribute to tallies for that simplex order, the total number of research articles  $P_{k}$  is conditioned by the simplex order $k$ under consideration. For this reason, curves for larger $k$ are shifted towards smaller research article count values, since the total number of MeSH per article is variable and generally increasing over the sample period, see {\bf Fig. \ref{fig2}}(B). 

Aside from this variation, the growth curves are largely consistent, each exhibiting an early period of relatively fast growth that declines in the early 2000s. 
Growth before and after this kink are best-fit by a linear model, indicating that there is an overall constant rate at which new simplices emerge that is in proportion to the rate of knowledge production, consistent with the results for patent IPC classes \cite{youn2015invention} and scientific article title-word combinations \cite{Fortunato_2017_Science}. 
Fitting the curves  with a linear model ($Y = A + \beta X$) over the more recent period 2005 to 2018, we obtain the slope $\beta$ corresponding to the average number of new simplices per article: $\beta_{k=1,Major}= 0.7$ and $\beta_{k=1,All}= 1.4$ pairs per article; $\beta_{k=2,Major}= 4.7$ and $\beta_{k=2,All}= 52$ triads per article;  $\beta_{k=3,Major}= 8$ and $\beta_{k=3,All}= 418$ quartets per article. While this latter number may seem unreasonably high, consider that there are ${12 \choose 4} = 495$ different combinations of 4 items selected from 12 items, and so any article with more than 12 MeSH might easily contribute 100s of new combinations to $C_{t}(S_{4})$.

All curves vary according to a systematic offset in both directions;  the difference in $C_{t}(S_{k})$ values is smaller for $k=3$ relative to $k=2$ than for $k=2$ relative to $k=1$, meaning that there is diminishing marginal increase in $C_{t}(S_{k})$ with increasing $k$. In other words, despite there being more possible combinations to tally for increasing $k$, fewer and fewer of these combinations appear to be realized. We return to the measurement of  the coverage of the combinatorial space in   Section \ref{Sec:Densification}. 

\vspace{-0.2in}
\subsubsection*{Combinatorial innovation and new knowledge}   
\vspace{-0.2in}
\noindent {\bf Figure \ref{fig2}}(B)  shows the total number of MeSH simplices of order $k$, denoted by $C_{t}(S_{k})$, as a function of the total number of MeSH ever used thru the same year.  Knowledge network growth parameterized according to vocabulary size are  consistent with the results of the previous section,  in that the vertical separation also  decreases with increasing $k$. In contradistinction to {\bf Figure \ref{fig2}}(A), there is no prominent kink in the empirical curves. As such,  an exponential growth model ($Y = A \exp[\gamma X]$) provides a  consistent fit over the entire range. We  plot  the best exponential  fit for each empirical curve,   with  $100\gamma$ corresponding to the  percent increase in $C_{t}(S_{k})$ per new MeSH term (i.e., those MeSH appearing for the first time that year): $100\gamma_{k=1,Major}= 1.6 \times 10^{-2}$ and $100\gamma_{k=1,All}= 1.9 \times 10^{-2}$ percent increase in  total pairs per new MeSH; $100\gamma_{k=2,Major}= 1.5 \times 10^{-2}$ and $100\gamma_{k=2,All}= 2.5 \times 10^{-2}$ percent increase in total triads per new MeSH;  $100\gamma_{k=3,Major}= 8.6 \times 10^{-3}$ and $100\gamma_{k=3,All}= 2.7 \times 10^{-2}$ percent increase in total quartets per new MeSH.  These values are less than unity, indicating a decreasing marginal increase in $C_{t}(S_{k})$ with each new MeSH, in analogy to the decreasing need for new words observed in a historical analysis of written language \cite{petersen2012languages}.

It is also worth  noting   that there is typically just an order of magnitude or less difference  between the curves for Major and All for a given $k$. 
This means that the bulk of $C_{t}(S_{k})$  are explained by combinations among research articles' primary concepts.
Yet the gap between  $C_{t,All}(S_{k})$ and $C_{t,Major}(S_{k})$ does appear to be widening over time for each $k$, and so the role of peripheral-mediated recombinant innovation does appear to become increasingly relevant. 

\begin{figure}
\centering{\includegraphics[width=0.99\textwidth]{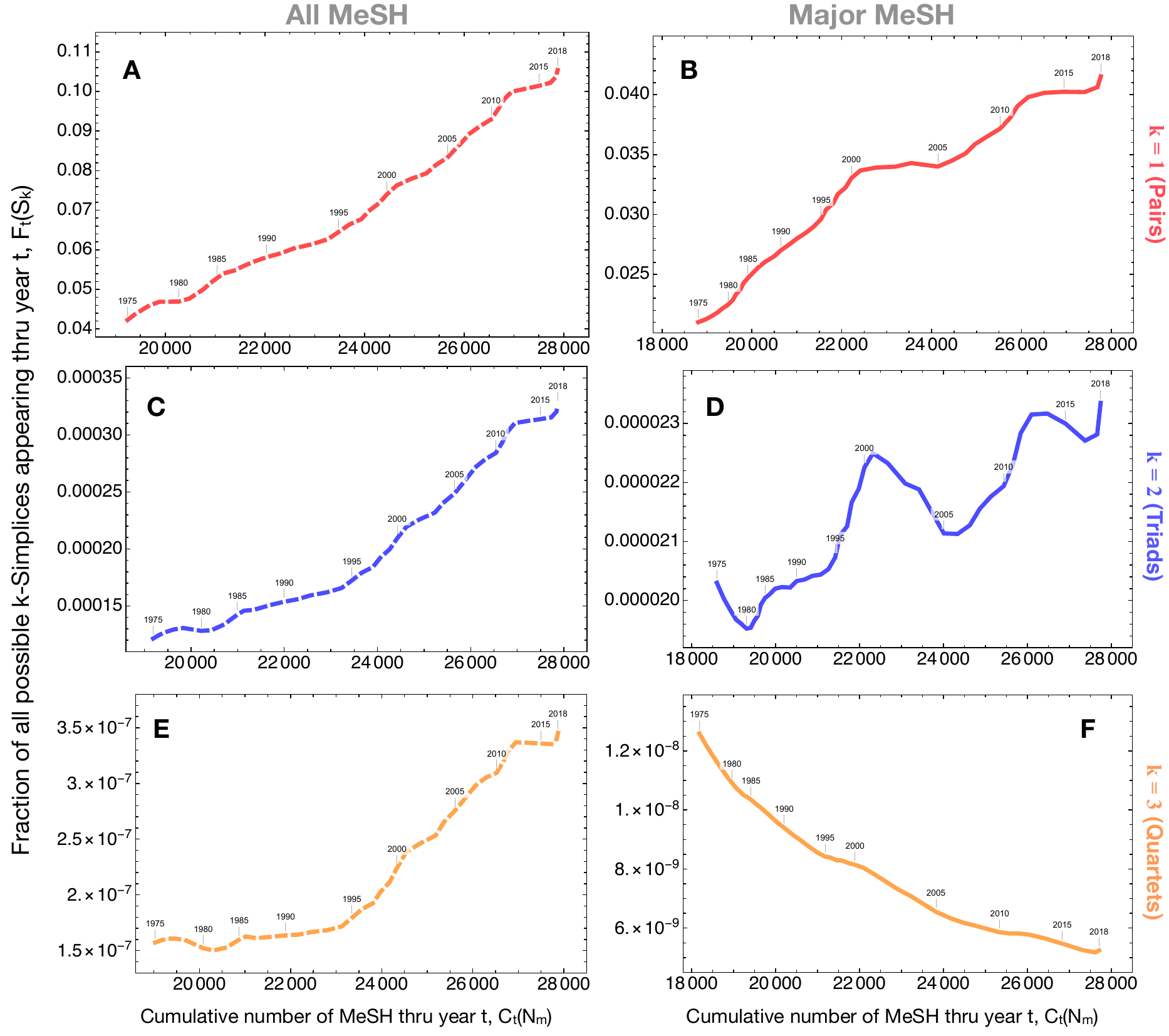}}
\caption{  \label{fig3}    {\bf Coverage of the combinatorial knowledge network.} 
 $F_{t}(S_{k})$ is the fraction  of all possible k-simplices  that are realized in research thru year $t$, plotted as a function of the size of the  MeSH vocabulary and for varying $k$. The panels on the left column (respectively, right column) show data calculated using All MeSH (resp., only  Major MeSH). For both MeSH refinements,  the combinatorial knowledge network shifts  from becoming more dense to more diffuse as $k=1$ to $k=3$, manifesting in increasingly smaller $F_{t}(S_{k})$ levels. However, the behavior of $F_{t}(S_{k})$ as a function of increasing vocabulary size $C_{t}(N_{m})$ does depend on the MeSH refinement. 
  {\bf (A,C,E)} The combinatorial knowledge network constructed from All MeSH exhibits uniform densification for all $k$. Notably, $F_{t}(S_{k})$ doubles  over the sample period in each case, with densification generally increasing from  1994 onward.
 {\bf (B,D,F)} As  $k$ increases the combinatorial knowledge network constructed from Major MeSH shifts from densification to diffusication as  $C_{t}(N_{m})$ increases. The $k=2$  scenario represents the margin for this behavior, as the amplitude of $F_{t}(S_{k})$ varies relatively little in this case; lacking an overall trend, this scenario   highlights 1994-2001 and 2007-2013 as periods with relatively high rates of recombinant innovation  per  new MeSH term.}
\end{figure}

\vspace{-0.2in}
\subsubsection*{Dynamics of combinatorial coverage -- Densification or Diffusication?} 
\label{Sec:Densification}
\vspace{-0.2in}
\noindent  Regarding the total realm of possibilities -- how much  has been explored by researchers over time? While the quantity  $C_{t}(N_{m})$ measures the growth of the revealed space of MeSH combinations in an absolute sense,  it does not account for all the combinations that have not been realized. The total number of k-simplex combinations possible  can be calculated exactly using the binomial coefficient ${C_{t}(N_{m}) \choose k}$, which grows as $n^{k}/k! \sim n^{k}$ with $n=C_{t}(N_{m})$. By way of example, since the total number of MeSH realized through 2018 is $C_{2018}(N_{m})$ = 27,875, then the total number of possible k-simplices are: ${27,875 \choose 2} = $ 388,493,875 $\sim 10^{8}$ pairs; ${27,875 \choose 3} =$ 3,609,496,592,625 $\sim 10^{12}$ triads; and ${27,875 \choose 4} =$ 25,150,972,257,411,000 $\sim 10^{16}$ quartets. Of course, not all of these combinations have been realized, nor do they merit exploring, and so the real question is  what proportion of the combinatorial space has been sampled over time.
To this end,  {\bf Fig. \ref{fig3}}  shows the fraction $F_{t}(S_{k}) = C_{t}(S_{k}) /  {C_{t}(N_{m}) \choose k}$ of the total space of combinations realized in research as a function of  the growing   MeSH vocabulary size.  

Focusing first on the calculating of coverage based upon the representation of the combinatorial knowledge network calculated using All MeSH, the main pattern is the strong and persistent growth in $F_{t}(S_{k})$ across each $k$, which more than doubles in each case. Hence, research is increasingly covering all knowledge combinations and so the knowledge network is densifying, even after accounting for  the increasing volume of knowledge. The rate of densification  appears to have increased since the mid 1990s.

An alternative perspective is offered by  focusing on just  Major MeSH, highlighted by  two prominent differences  across the curves for varying $k$. First, the baseline levels of $F_{t}(S_{k})$ decrease rapidly with increasing $k$, meaning that less and less of the combinatorial space has been covered when considering higher-dimensional representations of knowledge recombination (this is also the case for All MeSH). To be specific, roughly 4\% of all Major MeSH pairs have been combined thru 2018, whereas only 6  per billion possible quartets  have been  revealed thru the same year.  Second, there is a systematic shift from  a $F_{t}(S_{1})$ that increases with time, to   $F_{t}(S_{3})$ that decreases with time. The $F_{t}(S_{2})$ curve increases marginally over the sample period, and so the sudden bursts during 1994-2001 and 2007-2013 are more visible (although trend deviations are also visible in the $k=1$ and $k=3$ curves as well), and reflect the  high rates of recombinant innovation per new MeSH term during these periods.  

\vspace{-0.2in}
\subsection*{Distinguishing two modes of innovation: Conceptual and Recombinant}
\vspace{-0.2in}
\subsubsection*{Conceptual innovation and growth of the  knowledge network}
\begin{figure}[h!]
\includegraphics[width=0.99\textwidth]{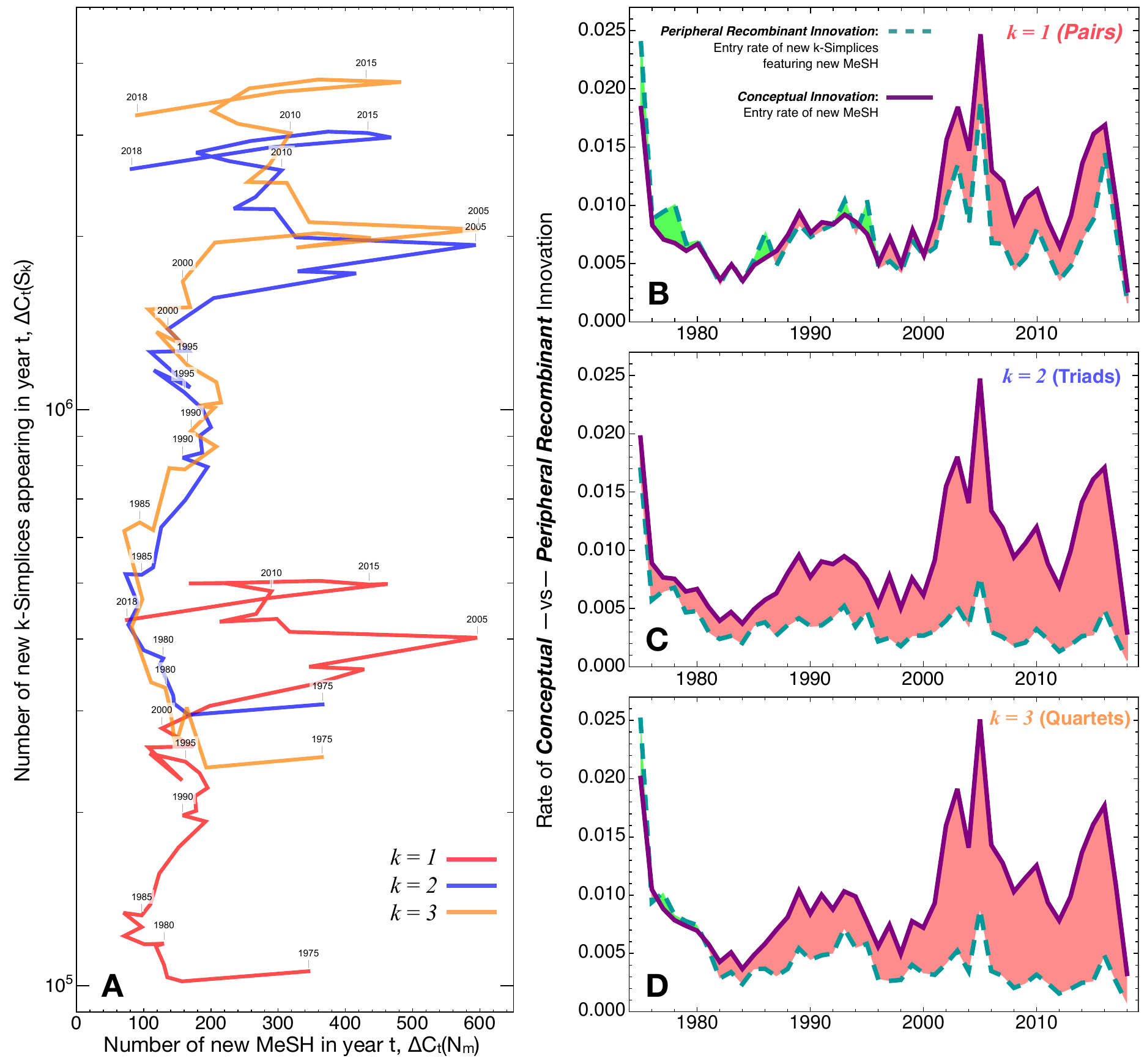}
\caption{  \label{fig4}    {\bf The dominant mode of recombinant innovation is  novel reconfigurations of pre-existing core knowledge.}   {\bf (A)}  Weak  relation between the emergence of new knowledge (proxied by rate of new MeSH) and the emergence of new recombinant knowledge associated with all new k-simplex configurations tallied in a given year. The burstiness of the relationship between new concepts and  recombinations may indicate a significant lag between new  knowledge  and downstream integration  into knowledge recombination. {\bf (B-D)} Measurement of combinatorial knowledge production associated with two innovation modes illustrated in {\bf Fig. \ref{fig1}}(E-G): (a) new k-simplices arising from the introduction of new MeSH (peripheral recombinant innovation), or (b) new k-simplices comprised of  pre-existing MeSH only (core recombinant innovation).
Plotted  in panels B, C, D  is the peripheral innovation rate $ r_{p}(t)= \Delta C_{t,\tt{Including New MeSH}}(S_{k})/\Delta C_{t}(S_{k})$ (dashed   curves) corresponding to mode $(a)$ (i.e., the frequency of new k-simplices incorporating new MeSH, as a proportion of all new k-simplices identified in a given year; note that   a complementary  frequency $r_{p}(t)$ corresponding to  mode (b) is very close to unity). For visual  comparison, solid curves represent the rate of conceptual innovation, $r_{m}(t) =  \Delta C_{t}(N_{m}) / C_{t}(N_{m})$, measured as the percent increase in the total number of MeSH. 
 In most years, the rate of peripheral recombinant innovation (a) is highly correlated with, and typically less than, the rate of conceptual innovation.  Moreover, the difference between the peripheral  and  conceptual rate of innovation  is increasing. The common patterns independent of $k$ suggest that higher order representations of combinatorial knowledge do not capture new phenomena, indicating that the traditional conceptualization as a knowledge network comprised of 1-simplces (links) is sufficient for understanding biomedical innovation dynamics. Data shown are calculated using only the Major MeSH terms.
 }
\end{figure}

\vspace{-0.2in}
\noindent  {\bf Figure \ref{fig4}}(A) shows the bursty  relationship between number of new k-Simplices appearing in a given year, denoted by $\Delta C_{t}(S_{k})$,  coinciding with the number of new MeSH occurring in the same year. While it would be easy to naively assume that a large number of new MeSH in a given year would result in a burst of new combinations, as a counterexample there are several notable periods with relatively few new MeSH entering the vocabulary, and relatively large increases in $\Delta C_{t}(S_{k})$. The entry of new MeSH is a proxy for conceptual innovation, and the lack of a strong relationship likely reflects the relatively slow rate of   diffusion through the combinatorial knowledge network. Hence, a future avenue of research could incorporate a larger  time-window for  $\Delta C_{t}(S_{k})$ to the time-scale of this combinatorial diffusion.

The burstiness of the relationship is common in magnitude and overall timing across the $k=$1,2,3 curves. Of particular note are the $k=2$ and $k=3$ curves which 
are quite similar in shape, which indicates that the $k=4$ representation provides no significant new insights (a conclusion that can also be drawn from  {\bf Fig. \ref{fig2}}).  Hence, the additional information contained in   higher-order representations appears to be marginal, providing additional support for measurement  frameworks based upon pairwise combinations  \cite{merchan2016sufficiency}. As such, the $k=2$ and $k=3$ representations of the  knowledge network appear  sufficient to capture the essential dynamics of combinatorial innovation as measured here.

\vspace{-0.2in}
\subsubsection*{Increasing concentration of conceptual innovation within single research articles}
\vspace{-0.2in}
\noindent As illustrated in {\bf Fig. \ref{fig2}}(E,F), conceptual innovation deriving from the birth  of new knowledge  gives rise to many new possible configurations. Yet there are so many missing configurations among already existing entities, that this alternative channel serves  as a very deep well providing many new recombinant options. So which innovation mode dominates? 
To address this question, we separate the new combinations in a each year, $\Delta C_{t}(S_{k})$, into two subsets: those that feature a new MeSH term (corresponding to peripheral recombinant innovation), and those that feature pre-existing MeSH (corresponding to core recombinant innovation). We calculate the  rate  of peripheral recombinant innovation   $ r_{p}(t)= \Delta C_{t,\tt{Including New MeSH}}(S_{k})/\Delta C_{t}(S_{k})$; the rate of core recombinant innovation is the complement, $r_{c}(t)=1-r_{p}(t)$.   Naturally, the baseline for comparison is the entry rate of new MeSH, measured in relative terms is given by $r_{m}(t)=\Delta C_{t}(N_{m})/ C_{t}(N_{m})$. 

{\bf Figure \ref{fig4}}(B) compares $r_{m}(t)$ and $r_{p}(t)$ for $k=$ 1,2,3, calculated for Major MeSH. Clearly, and not surprisingly, the dominant proportion of recombinant innovation is among existing primary knowledge, $r_{c}(t)\gg r_{p}(t)$. More interesting is the increasing gap between  $r_{m}(t)$ and $r_{p}(t)$, meaning that peripheral recombinant innovation is increasingly concentrated in relatively fewer k-simplices. By way of counterexample, if new knowledge did not tend to cluster in new k-simplices, then the rates $r_{m}(t)$ and $r_{p}(t)$ would be nearly identical, notwithstanding random fluctuations.  Because each k-simplex is a higher-order representation of the conceptual dimension of a single research article, this pattern means that conceptual innovation is increasingly concentrating in single research articles.

\vspace{-0.3in}
\section*{Discussion}
\vspace{-0.2in}
\noindent In this work we present a framework for systematically representing higher-order combinations of research article topics as a tool for measuring and better understanding the role of combinatorial innovation in science. We leverage the Medical Subject Heading (MeSH) ontology of keywords that are uniformly implemented within PubMed, are constantly being updated, and consisting in the present study of 27,875 individual keywords spanning a wide range of biological, medical, chemical, ecological concepts as well as equipment and other techno-informatic entities and methods \cite{BMConvergence_2021}. The excellent coverage and consistency of this ontology implemented within PubMed facilitates historical analysis to the extent that we can measure with certainty the year in which a new concept and new combination of concepts first appeared in the  literature.

In an effort to  systematize our approach, we adopted a  framework for counting all $k$-simplex variants of order $k=$1 (corresponding to MeSH pairs), 2 (triads) and 3 (quartets) in order to explore whether unexpected patterns of combinatorial innovation emerged at higher-order representations (see {\bf Fig. \ref{fig1}}). For the most part, we did not uncover information at $k=$ 3 that was not revealed at $k=$ 1. Nevertheless, three orders of $k$ were needed in order to differentiate spurious differences from trends. Consistent with prior research, we conclude that new combinations emerge roughly at a constant rate \cite{youn2015invention,Fortunato_2017_Science} (see {\bf Fig. \ref{fig2}}).

In addition to varying $k$, we also explored  differences based upon the two-level MeSH assignment, where Major MeSH represent primary research topics and entities, and the remaining MeSH represent more peripheral elements that were nevertheless integral to the research. Results using Major and All MeSH to measure the growth of the combinatorial knowledge network were overall consistent (see {\bf Fig. \ref{fig2}}). The exception being the analysis of the fraction of  MeSH combinations that  have been realized, which did yield fundamental differences between the representations based upon just Major versus All MeSH  (see {\bf  Fig. \ref{fig3}}). In particular, for All MeSH we observe a consistent densification for all $k$, whereas for Major MeSH we observed densification for $k=1$, marginal densification for $k=2$, and diffusication for $k=3$. Since the knowledge network constructed from Major MeSH represents the backbone of scientific concepts, this result indicates that most combinations are sampled  from a primary  set of conventional  topics, consistent with prior research analyzing  the co-occurrence of articles cited in reference lists \cite{Uzzi:2013}.  

Another phenomena highlighted  in {\bf Fig. \ref{fig3}}(D) is the identification of 1994-2001 and 2007-2013 as periods with relatively high rates of recombinant innovation per new MeSH term.
The  burst during 1994-2001 coincides with the  genomics revolution and the culminating success of the Human Genome Project   \cite{Petersen:2018}, a period featuring  deep convergence of technological applications and informatic methods from computer science to transform and accelerate the capabilities, scale and scope of biological sciences \cite{BMConvergence_2021}. In a similar fashion, the second innovation burst coincides with the continuation of transdisciplinary convergence efforts worldwide, many driven by national funding initiatives aiming to harness convergence,  as exemplified by the emergence of Human Brain projects in the US, Europe and Asia aiming to transform our understanding of the structure-function problem as it relates to understanding complexity, addressing the global burden of mental health problems, and the understanding and development of   artificial intelligence  \cite{sharp2011promoting,HBP_2020}.

Another objective in this work was to compare the rates of conceptual (i.e., the entry rate of new concepts) and recombinant innovation (i.e, the entry rate of new combinations) (see {\bf Fig. \ref{fig4}}), which confirms that the rate of recombinant innovation is overwhelmingly dominant. Because this disparity was largely anticipated, reflecting the sheer size of the MeSH knowledge corpus and the vast number of possible combinations, we instead focused on two types of recombinant innovation -- those new combinations that combine only pre-existing concepts (core) and those that incorporate new concepts (peripheral).
Results show that 99\% of new combinations did not incorporate new concepts, with the dynamics of  peripheral recombinant innovation closely matching (albeit typically lower) than the rate of conceptual innovation. Yet  comparing the difference between these two fundamental innovation rates for increasing $k$ revealed a widening gap, meaning that with each new piece of knowledge, there is a decreasing relative impact on the connectivity of the combinatorial knowledge network.  
Another interpretation of this trend considers the rapid integration of computational methods within the domain of biomedical science \cite{BMConvergence_2021}, such that innovative research increasingly incorporates multiple innovations simultaneously, e.g. a new algorithm or  instrument facilitating a new insight about a new biological process, such that neither would have happened without the demand for the other. Such ``combination reactions'' are the analog of chemical reactions requiring multiple reactants, and provide support for a triple-helix model of biomedical innovation in which technology plays an increasingly important role as catalyzer \cite{petersen2016triple,Petersen:2018}.  Indeed, the accelerating innovation by tapping  higher-order interactions is  a  principle value propositions of {\it convergence science},  which ascribes to the coming together of originally distinct fields and sectors \cite{Roco:2013,NRC:2014,Petersen:2018,HBP_2020,ConvergenceMethods_2022}, and thus represents a more macro-level representation of combinatorial innovation. \\

\noindent {\bf Acknowledgments} AMP acknowledges financial support from a Hellman Fellow award that was critical to completing this project. Publication and MeSH keyword ontology data are openly available from \href{https://pubmed.ncbi.nlm.nih.gov/}{https://pubmed.ncbi.nlm.nih.gov/}.

\clearpage
\newpage

\bibliographystyle{pnas.bst}
\bibliography{CombInnov}

\begin{thebibliography}{10}

\bibitem{BMConvergence_2021}
Yang D, Pavlidis I, Petersen AM
\newblock (2021) Biomedical convergence facilitated by the emergence of
  technological and informatic capabilities.
\newblock \emph{ArXiv e-print: 2103.10641}.

\bibitem{roehner2002patterns}
Roehner BM
\newblock (2002) \emph{Patterns of speculation: a study in observational
  econophysics}, See Table 1.1b, pp. 13
\newblock (Cambridge University Press).

\bibitem{bonaccorsi2008search}
Bonaccorsi A
\newblock (2008) Search regimes and the industrial dynamics of science.
\newblock \emph{Minerva} 46:285.

\bibitem{helbing2013globally}
Helbing D
\newblock (2013) Globally networked risks and how to respond.
\newblock \emph{Nature} 497:51--59.

\bibitem{kauffman1995home}
Kauffman S
\newblock (1995) \emph{At home in the universe: The search for laws of
  self-organization and complexity}
\newblock (Oxford University Press, USA).

\bibitem{thurner2018introduction}
Thurner S, Hanel R, Klimek P
\newblock (2018) \emph{Introduction to the theory of complex systems}
\newblock (Oxford University Press).

\bibitem{sadybekov2022synthon}
Sadybekov AA, {et~al.}
\newblock (2022) Synthon-based ligand discovery in virtual libraries of over 11
  billion compounds.
\newblock \emph{Nature} 601:452--459.

\bibitem{tshitoyan2019unsupervised}
Tshitoyan V, {et~al.}
\newblock (2019) Unsupervised word embeddings capture latent knowledge from
  materials science literature.
\newblock \emph{Nature} 571:95--98.

\bibitem{HBP_2020}
Petersen AM, Ahmed ME, Pavlidis I
\newblock (2021) Grand challenges and emergent modes of convergence science.
\newblock \emph{{Humanities and Social Sciences Communications}} 8:194.

\bibitem{arroyave2021social}
Arroyave FJ, {et~al.}
\newblock (2021) On the social and cognitive dimensions of wicked environmental
  problems characterized by conceptual and solution uncertainty.
\newblock \emph{Advances in Complex Systems} 24.

\bibitem{ConvergenceMethods_2022}
Petersen AM, Arroyave F, Pavlidis I
\newblock (2022) Methods for measuring social and conceptual dimensions of
  convergence science.
\newblock \emph{Under Review -- SSRN e-print: 4117933} pp 1--11.

\bibitem{Leahey_Sociological_2014}
Leahey E, Moody J
\newblock (2014) Sociological innovation through subfield integration.
\newblock \emph{Social Currents} 1:228--256.

\bibitem{petersen2016triple}
Petersen AM, Rotolo D, Leydesdorff L
\newblock (2016) {A Triple Helix Model of Medical Innovation: Supply, Demand,
  and Technological Capabilities in terms of Medical Subject Headings}.
\newblock \emph{Research Policy} 45:666--681.

\bibitem{mishra2016quantifying}
Mishra S, Torvik VI
\newblock (2016) \emph{Quantifying conceptual novelty in the biomedical
  literature}
\newblock Vol.{}~22, pp 468--472.

\bibitem{petersen2018mobility}
Petersen AM
\newblock (2018) Multiscale impact of researcher mobility.
\newblock \emph{Journal of The Royal Society Interface} 15:20180580.

\bibitem{gerlach2018network}
Gerlach M, Peixoto TP, Altmann EG
\newblock (2018) A network approach to topic models.
\newblock \emph{{Science Advances}} 4:eaaq1360.

\bibitem{palchykov2021network}
Palchykov V, Krasnytska M, Mryglod O, Holovatch Y
\newblock (2021) Network of scientific concepts: empirical analysis and
  modeling.
\newblock \emph{Advances in Complex Systems} 24:214001.

\bibitem{arroyave2022NPS}
Arroyave FJ, Jenkins J, Shackleton S, Petersen AM
\newblock (2022) {Research alignment in the U.S. National Park Service: Impact
  of transformative science policy on the supply of scientific knowledge for
  protected area management}.
\newblock \emph{SSRN e-print: 4044154} pp 1--28.

\bibitem{Uzzi:2013}
Uzzi B, Mukherjee S, Stringer M, Jones B
\newblock (2013) Atypical combinations and scientific impact.
\newblock \emph{Science} 342:468--472.

\bibitem{MESH}
{MeSH}
\newblock (Accessed 3/2020) {The MeSH (Medical Subject Headings) system: A
  controlled vocabulary thesaurus used for indexing PubMed articles}.
  (\url{http://www.ncbi.nlm.nih.gov/mesh}).

\bibitem{MESHMap}
Leydesdorff L, Rotolo D, Rafols I
\newblock (2012) Bibliometric perspectives on medical innovation using the
  medical subject headings of pub med.
\newblock \emph{Journal of the American Society for Information Science and
  Technology} 63:2239--2253.

\bibitem{rodriguez2013quantifying}
Rodriguez-Esteban R, Loging WT
\newblock (2013) Quantifying the complexity of medical research.
\newblock \emph{Bioinformatics} 29:2918--2924.

\bibitem{shi2015weaving}
Shi F, Foster JG, Evans JA
\newblock (2015) Weaving the fabric of science: Dynamic network models of
  science's unfolding structure.
\newblock \emph{Social Networks} 43:73--85.

\bibitem{youn2015invention}
Youn H, Strumsky D, Bettencourt LM, Lobo J
\newblock (2015) Invention as a combinatorial process: evidence from us
  patents.
\newblock \emph{Journal of the Royal Society interface} 12:20150272.

\bibitem{Fortunato_2017_Science}
Fortunato S, {et~al.}
\newblock (2018) {Science of Science}.
\newblock \emph{Science} 359:eaao0185.

\bibitem{weitzman1998recombinant}
Weitzman ML
\newblock (1998) Recombinant growth.
\newblock \emph{The Quarterly Journal of Economics} 113:331--360.

\bibitem{fleming2001recombinant}
Fleming L
\newblock (2001) Recombinant uncertainty in technological search.
\newblock \emph{{Management Science}} 47:117--132.

\bibitem{napolitano2018technology}
Napolitano L, Evangelou E, Pugliese E, Zeppini P, Room G
\newblock (2018) Technology networks: the autocatalytic origins of innovation.
\newblock \emph{Royal Society open science} 5:172445.

\bibitem{petersen_citationinflation_2018}
Petersen AM, Pan RK, Pammolli F, Fortunato S
\newblock (2018) Methods to account for citation inflation in research
  evaluation.
\newblock \emph{Research Policy} 48:1855--1865.

\bibitem{battiston2020networks}
Battiston F, {et~al.}
\newblock (2020) Networks beyond pairwise interactions: structure and dynamics.
\newblock \emph{Physics Reports} 874:1--92.

\bibitem{battiston2021physics}
Battiston F, {et~al.}
\newblock (2021) The physics of higher-order interactions in complex systems.
\newblock \emph{Nature Physics} 17:1093--1098.

\bibitem{merchan2016sufficiency}
Merchan L, Nemenman I
\newblock (2016) On the sufficiency of pairwise interactions in maximum entropy
  models of networks.
\newblock \emph{Journal of Statistical Physics} 162:1294--1308.

\bibitem{feynman1949space}
Feynman RP
\newblock (1949) Space-time approach to quantum electrodynamics.
\newblock \emph{Physical Review} 76:769.

\bibitem{petersen2012languages}
Petersen AM, Tenenbaum JN, Havlin S, Stanley HE, Perc M
\newblock (2012) Languages cool as they expand: Allometric scaling and the
  decreasing need for new words.
\newblock \emph{Scientific Reports} 2:943.

\bibitem{Petersen:2018}
Petersen AM, Majeti D, Kwon K, Ahmed ME, Pavlidis I
\newblock (2018) Cross-disciplinary evolution of the genomics revolution.
\newblock \emph{Science Advances} 4:eaat4211.

\bibitem{sharp2011promoting}
Sharp PA, Langer R
\newblock (2011) Promoting convergence in biomedical science.
\newblock \emph{Science} 333:527--527.

\bibitem{Roco:2013}
Roco M, Bainbridge W, Tonn B, Whitesides G
\newblock (2013) \emph{Converging Knowledge, Technology, and Society: Beyond
  Convergence of Nano-Bio-Info-Cognitive Technologies}
\newblock (Springer, New York).

\bibitem{NRC:2014}
{National Research Council}
\newblock (2014) \emph{Convergence: Facilitating transdisciplinary integration
  of life sciences, physical sciences, engineering, and beyond}
\newblock (National Academies Press, Washington, D.C.).

\end{thebibliography}

\end{document}